\documentclass[aps,prl,preprintnumbers,showpacs,nofootinbib,floatfix,amsmath]{revtex4}
\usepackage{graphicx}
\usepackage{comment}
\usepackage{dcolumn}
\usepackage{bm}
\begin{document}
\newcommand{\newc}{\newcommand}
\newc{\ra}{\rightarrow}
\newc{\lra}{\leftrightarrow}
\newc{\beq}{\begin{equation}}
\newc{\eeq}{\end{equation}}
\newc{\barr}{\begin{eqnarray}}
\newc{\earr}{\end{eqnarray}}
\newcommand{\Od}{{\cal O}}
\newcommand{\lsim}   {\mathrel{\mathop{\kern 0pt \rlap
  {\raise.2ex\hbox{$<$}}}
  \lower.9ex\hbox{\kern-.190em $\sim$}}}
\newcommand{\gsim}   {\mathrel{\mathop{\kern 0pt \rlap
  {\raise.2ex\hbox{$>$}}}
  \lower.9ex\hbox{\kern-.190em $\sim$}}}
\title {EXPLORING NEW FEATURES OF NEUTRINO OSCILLATIONS WITH VERY LOW ENERGY
MONOENERGETIC NEUTRINOS}
%
%
%
%
%
%
\author{J.D. Vergados$^{1}$, and Yu.N. Novikov$^{2}$ }
%
\affiliation{
{\it 1 University of Ioannina, Ioannina, GR 45110, Greece
\\E-mail:Vergados@cc.uoi.gr}}
\affiliation{
{\it 2 Petersburg Nuclear Physics Institute, 188300, Gatchina,Russia}}
\begin{abstract}
In the present work we propose to study neutrino oscillations employing sources of monoenergetic neutrinos 
following electron capture by the nucleus. Since the neutrino energy is very low the smaller of the two oscillation lengths, $L_{23}$,
appearing in this electronic neutrino  disappearance experiment can be so small that the full oscillation can take place inside the detector and one may determine very
accurately the neutrino oscillation parameters.
 Since in this case the oscillation probability is proportional to $\sin^2{2 \theta_{13}}$,  one can  measure or set a better limit on the unknown parameter $\theta_{13}$. This is quite important, since, if this mixing angle vanishes, there is not going to be CP violation in the leptonic sector. The best way to detect it is by measuring electron recoils in neutrino-electron scattering. One, however, has to pay the price that  the expected counting rates are very small. Thus one needs a very intensive neutrino source and a large detector with as low as possible energy threshold and high energy and position resolution. Both spherical gaseous and cylindrical liquid detectors are studied. Different  source candidates are considered.
\end{abstract}
\date{\today}

\maketitle
\section{Introduction.}
The discovery of neutrino oscillations can be considered as one of the greatest triumphs of modern physics.
It began with atmospheric neutrino oscillations \cite{SUPERKAMIOKANDE}interpreted as
 $\nu_{\mu} \rightarrow \nu_{\tau}$ oscillations, as well as
 $\nu_e$ disappearance in solar neutrinos \cite{SOLAROSC}. These
 results have been recently confirmed by the KamLAND experiment \cite{KAMLAND},
 which exhibits evidence for reactor antineutrino disappearance.
  As a result of these experiments we have a pretty good idea of the neutrino
mixing matrix and the two independent quantities $\Delta m^2$, e.g $|m_2^2-m^2_1|$ and $|m^2_3-m^2_2|$.
 Fortunately these
two  $\Delta m^2$ values are vastly different, $$\Delta
m^2_{21}=|m_2^2-m_1^2|=(7.65^{+0.23}_{-0.20})\times 10^{-5}(eV)^2$$ and
$$\Delta m^2_{32}=|m_3^2-m_2^2|=(2.4^{+0.12}_{-0.11})\times 10^{-3}(eV)^2.$$
 This means that the relevant $L/E$ parameters are very different. Thus for a given energy the experimental results can approximately be described as two generation oscillations. For an accurate description  of the data, however, a three generation analysis  \cite{BAHCALL02},\cite{BARGER02} is necessary.

 In all of these analyses the
oscillation length is much larger than the size of the detector. So one is able to see the effect, if the detector is
placed in the right distance from the source. It is, however, possible to design an experiment with an oscillation
length of the order of the size of the detector. In this case one can start with zero oscillation near the
source, proceed to maximum oscillation near the middle of the detector and end up again with no oscillation on
the other end. This is achieved, if one considers a neutrino source with as low as practical  neutrino energy. The main requirements are as follows:
\begin{itemize}
\item The neutrinos should have as low as possible energy so that the oscillation length can
be minimized. At the same time it should not be too low, so that
the neutrino-electron cross section is sizable.
\item A monoenegetic source has the advantage that some of the features of the oscillation patterns are not washed out by the averaging over a continuous neutrino spectrum.
\item The life time of the source should be suitable for the experiment to be performed. If it is too short,
the time available will not be adequate for the execution of the experiment. If it is too long, the number
of counts during the data taking will be too small. Then one will face formidable backgrounds and/or large experimental uncertainties.
\item The source should be cheaply available in large quantities.
\end{itemize}
 Clearly a  compromise has to be made in the selection of the source.
 It appears, unfortunately, that  the oscillation length one may have to live with,
 may be quite a bit longer than that ofthe previously considered triton
 source with a maximum energy of $18.6$ keV, i.e.  an   average oscillation length is $6.5$m.

 At such low energies the only neutrino detector, which is  sensitive to neutrino oscillations, is one, which is capable of detecting recoiling electrons. With this detector one is sensitive to all neutrino flavors \cite{HOOFT}, \cite{REINES}. In fact one can detect:
\begin{itemize}
\item electrons which are produced by electron neutrinos via both the charged and the neutral current interaction. These
 will  manifest as {\bf electron neutrino disappearance}
\item electrons are produced from the other two neutrino flavors due to the neutral
current interaction. These two flavors are due to the  {\bf muon and tau neutrino appearance}.
\end{itemize}
 Since the importance of the
cross section due to the charged current relative to that due to the neutral current depends on the electron energy, one has a novel feature, i.e. the  effective oscillation probability
depends on the electron energy. Thus the results may appear as disappearance oscillation in some kinematical regime
and as appearance oscillation in some other regime. In this paper we will examine how these features can best be
exploited in determining the neutrino oscillation parameters.
\section{ Elastic Neutrino Electron Scattering}
 For low energy neutrinos the historic process neutrino-electron scattering \cite{HOOFT} \cite{REINES}
 is very useful.
The differential cross section \cite{VogEng} takes the form
\begin{equation}
\frac{d\sigma}{dT}=\left(\frac{d\sigma}{dT}\right)_{weak}+
\left(\frac{d\sigma}{dT}\right)_{EM} \label{elas1a}
\end{equation}
The second term, due to the neutrino magnetic moment, is inversely proportional to the electron
energy and, at the low electron energies of the present set up, may be used in improving
 the current limit of the neutrino magnetic moment by two orders of magnitude. It is not, however,
 the subject of the present study, which is concerned with neutrino oscillations. The cross section in the rest frame of the initial electron due to weak interaction alone becomes:
 \begin{eqnarray}
 \left(\frac{d\sigma}{dT}\right)_{weak}&=&\frac{G^2_F m_e}{2 \pi}
 [(g_V+g_A)^2\\
\nonumber
&+& (g_V-g_A)^2 [1-\frac{T}{E_{\nu}}]^2
+ (g_A^2-g_V^2)\frac{m_eT}{E^2_{\nu}}]
 \label{elasw}
  \end{eqnarray}
 $$g_V=2\sin^2\theta_W+1/2~~ (\nu_e)~,~g_V=2\sin^2\theta_W-1/2~~ (\nu_{\mu},\nu_{\tau})$$
 $$g_A=1/2~~~~ (\nu_e)~~~~,~~~~g_A=-1/2~~~~ (\nu_{\mu},\nu_{\tau})$$
 For antineutrinos $g_A\rightarrow-g_A$.
 
  The scale is set by the week interaction:
\beq \frac{G^2_F m_e}{2 \pi}=4.45\times 10^{-48}~\frac{cm^2}{keV}
\label{weekval}
\eeq
The above equation must be modified, if one wishes to include the fact that the initial electron is bound. For neutrino energies in the tens of keV, however, it has been shown by Gounaris, Paschos and Porfyriadis \cite{GPP04} that such a modification  lowers the cross section by no more than 10\%. So we are not going to consider such effects here.
\\The electron energy depends on the  neutrino energy and the
scattering angle and is given by:
\beq
T= \frac{2~m_e (E_{\nu}\cos{\theta})^2}{(m_e+E_\nu)^2-(E_{\nu}
\cos{\theta})^2}
\label{Te}
\eeq

\begin{figure}[!ht]
\begin{center}
\rotatebox{90}{\hspace{-0.0cm} {$\sigma (\nu_e,e^-)\rightarrow 10^{-46}cm^2$}}
\includegraphics[scale=1.0]{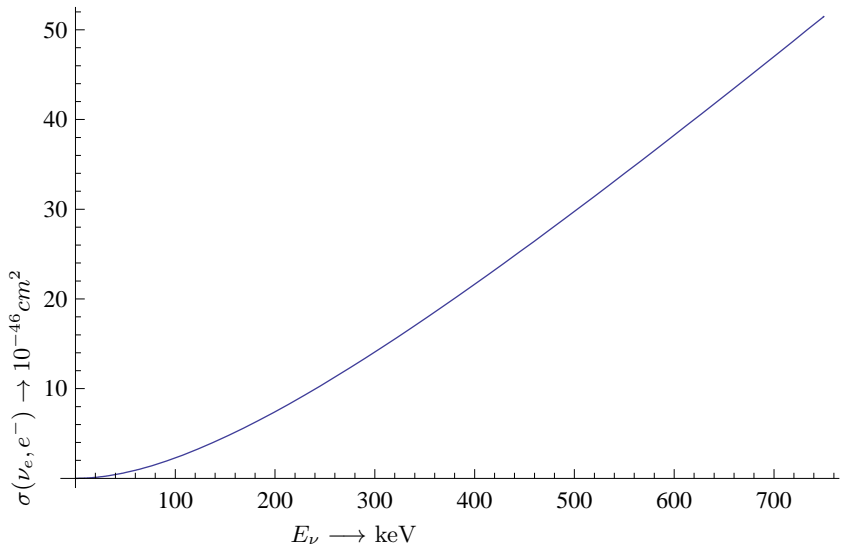}\\
{\hspace{-2.0cm} {$E_{\nu}\longrightarrow$ keV}}\\
 \caption{The  total $(\nu_e,e^-)$ cross section in the absence of oscillations as a function of the neutrino energy. In the present work the neutrinos have a definite energy.}
 \label{fig:sigmatotal}
  \end{center}
  \end{figure}

The last equation for sufficiently low energies can be simplified as follows:
 \beq
T \approx \frac{ 2(E_\nu \cos{\theta})^2}{m_e}
\label{Teap}
\eeq
 The maximum electron
energy depends on the neutrino energy squared.
The total cross section as a function of neutrino energy is shown in Fig. \ref{fig:sigmatotal}.
\section{ NEUTRINO SHORT BASELINE OSCILLATIONS}

The electron neutrino, produced in weak interactions, can be expressed in terms of the three mass eigenstates as follows:
\beq
\nu_e=\cos{\theta_{12}} \cos{\theta_{13}}~\nu_1+\sin{\theta_{12}} \cos{\theta_{13}}\,\nu_2+\sin{\theta_{13}}~ e^{i\delta} \nu_3
\label{nue}
\eeq
 Then the  $\nu_e$ disappearance oscillation probability is
given by:
\barr
 P(\nu_e \rightarrow \nu_e )&=&1- 
 [
\cos ^4 {\theta _{13}}   \sin ^2 {2\theta_{12}}
\sin^2 {(\pi \frac{L}{L_{21}})}
\nonumber\\
&+ & \sin ^2{\theta_{12}} \sin ^2{2\theta _{13}}
\sin^2{ (\pi \frac{L}{L_{32}})}
\nonumber\\
& &
+ \cos^2{\theta _{12}} \sin ^2{2 \theta _{13}}
 \sin^2 {(\pi \frac{L}{L_{31}})}
]
 \label{disap}
 \earr
with
\beq 
L_{ij}=\frac{4 \pi E_{\nu}}{m_i^2-m_j^2}
\label{OscLength}
\eeq
 $sin \theta_{13}$ is a small quantity constrained by the
CHOOZ experiment and it will be further investigated below. In the limit in which it is neglected, one recovers
the standard expression
familiar from the solar neutrino oscillation analysis, if we take
 $s_{12}\approx sin \theta_{solar}$ , $c_{12}\approx cos \theta_{solar}$, $\theta_{solar}$ as determined from the
solar neutrino data.
 Assuming $\Delta m^2_{31} \approx \Delta m^2_{32}$ we find:
\barr
P(\nu_e \rightarrow \nu_e)\approx
1-&&\large [(\sin 2 \theta_{solar})^2 \sin^2(\pi \frac{L}{L_{21}})+\nonumber\\
 && \sin^2(2 \theta_{13}) \sin^2(\pi \frac{L}{L_{32}}) \large]
 \label{pdisap}
 \earr
where the term  proportional to $\sin^2{(2\theta_{13})}$, connected with the small oscillation length \cite{NOSTOS1}
$L_{32}$, ($L_{21}\approx 32 L_{32}$), is relevant for the experimental approaches discussed in the present paper.
As we have already mentioned in connection with the NOSTOS experiment\cite{NOSTOS1}, in addition to the electronic neutrinos, which are depleted by the oscillations,
 is sensitive to the other two neutrino flavors, which can also produce electrons
 via the neutral current interaction. These flavors are
  generated via the appearance oscillation:
\barr
P(\nu_e \rightarrow \sum_{\alpha \ne e} \nu_{\alpha})\approx&&
(\sin 2 \theta_{solar})^2 \sin^2(\pi \frac{L}{L_{21}})+\nonumber\\
  &&\sin^2{(2\theta_{13})} \sin^2(\pi \frac{L}{L_{32}})
 \earr
 Thus the number of the scattered electrons, which bear this rather unusual
oscillation pattern, is proportional to
 the $(\nu_e,e^-)$ scattering cross section
with a proportionality constant $C(E_{\nu},T)$ given by:
\barr
C(E_{\nu},T)&=& 1-\chi(E_{\nu},T)
\nonumber\\
& & (\sin 2 \theta_{solar})^2 \sin^2(\pi \frac{L}{L_{21}})+
 \sin^2{(2\theta_{13})} \sin^2(\pi \frac{L}{L_{32}}).
\nonumber\\
\label{oscprob}
 \earr
The function $\chi(E_{\nu},T)$, which represents the relative difference between the cross section of the  electronic neutrino and its other two flavors, has been previously
discussed  \cite{NOSTOS1}.
\section{Geometric Considerations}
The total neutrino electron scattering cross section as a function of $x$ and $L$ can be cast in the form:
\beq
\sigma(L,x)=\sigma(0,x)\left( 1-\chi(x) p(L,x)\right )
\label{sigmatot1}
\eeq
with $x=\frac{E_{\nu}}{m_e}$ and
\beq
\sigma(0,x)=\frac{G^2_F m^2_e}{2 \pi} \frac{x^2 \left(17.7464 x^2+15.3098 x+3.36245\right)}{(2 x+1)^3}
\label{sigmatot2}
\eeq
is the total cross section in the absence of oscillations. Furthermore
\barr
p(L,x)&=&\sin ^2\left(\frac{ 0.0595922 L}{330x}\right) \sin ^2(2 \theta_{solar} )+\nonumber\\
& &\sin^2\left(\frac{0.0595922 L}{10x}\right) \sin ^2\left(2 \theta_{13}\right) 
 \earr
 with $L$ the source detector distance in meters and
\beq
\chi(x)=\frac{2.8664 x^2+4.1498 x+1.50245}{17.7464 x^2+15.3098 x+3.36245}
\eeq
 We will assume that the volume of the source is much smaller than the volume of the detector.

We will consider two possibilities:
\subsection{Spherical detector with the source at the origin}
The number of events between $L$ and $L+dL$ is given by:
\beq
dN=N_{\nu} n_e \frac{4 \pi L^2dL}{4 \pi L^2} \sigma(L,x)=N_{\nu} n_e dL \sigma(L,x) 
\eeq
or
\beq
 \frac{dN}{dL}=N_{\nu} n_e \sigma(L,x )
 \label{eventsph}
\eeq
To compare with other geometries we rewrite this as follows:
\beq
 \frac{dN}{d\rho}=R_0 N_{\nu} n_e \sigma(x,\rho)
 \eeq
 or
 \beq
  \frac{dN}{d\rho}=\frac{G^2_F m^2_e}{2 \pi} R_0 N_{\nu} n_e g_s(\rho)\tilde{\sigma}(x,\rho),~~g_s(\rho)=1
\eeq
with $N_{\nu}$ the number of neutrinos emitted by the source, $n_e $ the density of electrons in the target, $R_0$ the radius of the target and $\rho=L/R_0$. $\tilde{\sigma}(x,\rho)$ is the neutrino - electron cross section in units of 
${G^2_F m^2_e}/{2 \pi}$. The geometry factor $g_s(\rho)$  is in this case unity.
\subsection{Cylindrical detector with the source at the center of one of its basis}
 The number of events between $r$ and $r+dr$ and $z$ and $z+dz$ is now given by:
 \beq
 dN=N_{\nu} n_e \frac{ 2 \pi r dr dz}{4\pi (r^2+z^2)} \sigma(x,\sqrt{r^2+z^2})
 \eeq
 which yields
 \beq
 \frac{dN}{d \rho d \zeta}=N_{\nu} n_e R\frac{1}{2} \frac{u \rho}{\zeta^2+u^2 \rho^2} \sigma(x,\frac{R}{u}\sqrt{\zeta^2+u^2 \rho^2},
 \label{Eq:cylinder}
 \eeq
  where $R$ is the radius of the cylinder, $u=R/h$ ($h$ is the length of the cylinder), $\rho=r/R$ and $\zeta=z/h$. This can be written as 
\beq
\frac{dN}{d \rho d \zeta}=\Lambda g_c(\rho,\zeta,u,R)\frac{1}{2}\tilde{\sigma}(x,\frac{R}{u}\sqrt{\zeta^2+u^2 \rho^2}),
\eeq
\beq
g_c(\rho,\zeta,u,R)= \frac{u \rho}{\zeta^2+u^2 \rho^2} ~~,~~\Lambda=\frac{G^2_F m^2_e}{2 \pi} R N_{\nu} n_e
\label{gcLambda}
\eeq
 Note that the geometric factor $g_c(\rho,\zeta,u,R)$, which is absent in the spherical geometry, in this case is less than unity.
 
\begin{figure}[!ht]
 \begin{center}
\includegraphics[scale=1.0]{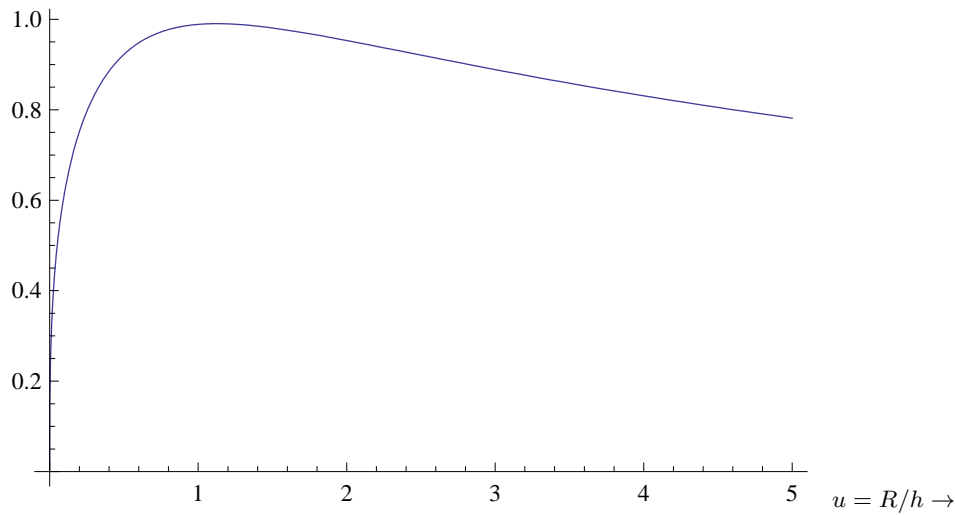}
\hspace*{-0.0cm} { $u=R/h\rightarrow$}
 \caption{The relative geometric merit, $N_c/N_s$, of a cylinder compared to that of a half sphere of the same volume as a function of $u=R/h$}
 \label{fig:geom}
 \end{center}
 \end{figure}

 It is instructive to compare the two geometries in the absence of oscillations. In such a case  we find that the number of events is given by:
 \beq
 N_c=N_{\nu} n_e R \sigma(x,0)\frac{1}{4} \left(2 \tan ^{-1}\left(\frac{1}{u}\right)+\frac{\log
   \left(u^2+1\right)}{u}\right)
 \eeq
 while for a half sphere of the same volume we find:
 \beq
  N_s=N_{\nu} n_e R \sigma(x,0)\frac{\sqrt[3]{\frac{3}{2}}}{2 \sqrt[3]{u}}
 \eeq
 The relative merit $N_c/N_s$ is shown in Fig. \ref{fig:geom}. This ratio becomes unity for $u=1.12$, which is much larger than the value of 11/91 of the planned detector. Note, however, that for u=11/91 the relative merit is 0.75, .i.e. not far from unity. 
\section{CANDIDATES FOR NEUTRINO OSCILLOMETRY}
One can ask whether the relevant candidates for small length oscillation measurements exist in reality. The analysis of known information on nuclides show that there are many cases with a small  neutrino energy in the orbital electron capture by nucleus. Since this process has the two-body mechanism, the total neutrino energy is equal to the difference of the total capture energy $Q_EC$ and binding energy of captured electron $B_i$ :
\beq
Q_{\epsilon}=Q_{EC}-B_i
\eeq                                                    
This value can be easily determined because the capture energies are usually known (or can be measured very precisely by the ion-trap spectrometry \cite{BNW10}) and the electron binding energies are tabulated \cite{LARKINS}. The main feature of the electron capture process is the monochromaticity of neutrino. This   paves the way for the neutrino oscillometry, which is introduced below. 
Table \ref{tab1} shows the relevant nuclides, which emit monoenrgetic neutrinos with energy less than 750 keV, half-life not more than $10^5$ y, and whose production rates in the neutron reactors can provide rather high intensity of neutrinos ($\ge 10^{12}$ s$^{-1}$).  Columns 2-4 of Table \ref{tab1} show the decay characteristics of the corresponding nuclides, column 5 gives a half value of the small oscillation length $L_{32}$, column 6 shows the recoil electron maximal energy in the neutrino scattering on electron, and column 7 presents the masses of samples that can reasonably be produced by irradiation of natural targets in reactors with the assumed neutron flux of $5\times10^{14 }$cm$^{-2}$s$^{-1}$ and known neutron capture cross-sections for individual nuclides. 
We can see from Table \ref{tab1} that there is a variety of nuclides with different half-lives and capture energies. Short half-life nuclides, as e.g. $^{51}$Cr, $^{55}$Fe,  $^{71}$Ge, and $^{103}$Pd, have an advantage in the sense that they can be rapidly produced in reactors, whereas long-lived ones, as $^{41}$Ca, $^{157}$Tb, $^{163}$Ho and $^{193}$Pt need long-term irradiation in order to accumulate necessary decay intensity. The nuclides  $^{157}$Tb and $^{163}$Ho can be accumulated after disintegration of $^{157}$Dy and $^{163}$Er, respectively, which in turn can be produced by neutron irradiation of natural dysprosium and erbium or enriched samples of $^{156}$Dy and $^{162}$Er. To produce a high intensity sources one needs  a huge amount of irradiated material. For the production of $^{71}$Ge with neutrino intensity of $2\times10^{18}$s$^{-1}$ (55 MCi)  the 300 kg of natural  Ge should be used with the irradiation time of 20 days at the reactor. Obviously, these irradiations in principle can be repeated many times. For the production of $^{51}$Cr  ca. 40 kg enriched target of $^{50}$Cr used for GALLEX-detector calibration \cite{HAMPEL} can be favorably borrowed. 
The long half-life implies the production of smaller source intensity even after a long-time neutron irradiation, which can not be fully compensated by subsequent long-time data acquisition. Meanwhile, just these long-lived nuclides with the corresponding small neutrino energies can provide as small as possible short oscillation lengths $L_{32}$. \\
The values of $L_{32}$  in the column 5 were determined by the formula of Eq. (\ref{OscLength}).
Thus we can write:
\beq
 L_{32} =  \frac{2.48[\mbox{m}] E_{\nu}}{\Delta m^2_{32}([\mbox{eV}])^2}
  \Rightarrow
 L_{32}[\mbox{m}]\approx E_{\nu} [\mbox{keV}]
 \label{L32}
\eeq                                                             
 The values in the square brackets in Eqs (\ref{L32}) indicate the dimensions used.
As can be seen from Table \ref{tab1} (column 5) one can roughly divide the nuclides presented there into two categories: those which have $L_{32}\le 50$ m and those with $L_{32} > 110$ m. 
For the former nuclides the TPC counting method can be used in the gas-filled NOSTOS sphere approach \cite{NOSTOS1}, \cite{Giomataris}, whereas for both and mainly for the latter category 
with the larger $L_{32}$, the long liquid scintillator (LS) detector \cite{OBERAUER} is more preferable. Both methods are discussed in a forthcoming section.
The goal of both approaches is to scan the monoenergetic neutrino-electron scattering events by measuring the electron recoil counts in a function of distance from the neutrino source prepared in advance at the reactor/s. This scan means point-by-point determination of scattering events along the detector dimensions within its position resolution. In the best cases these events can be observed as a smooth curve which reproduces the neutrino disappearance probability. We call this measurement a "neutrino oscillometry". It is worthwhile to note again that the oscillometry is suitable for monoenergetic neutrino since, then, one deals with a  single oscillation length $L_{32}$. This is obviously not a case for antineutrino, since, in this instance, one extracts only an effective oscillation length. Thus some information may be lost due to the folding with the continuous neutrino energy spectrum. 

\begin{table}[t]
\caption{Possible candidates for neutrino box-oscillometry. Only the nuclides with half-lives shorter than 10$^5$  y and with the neutrino production intensity higher than 10$^{12}$ s$^{-1}$ have been chosen.}
\label{tab1}
\begin{center}
\begin{tabular}{|c|c|c|c|c|c|c|c|}
\hline
\hline
 &   &  &  & & & & \\
Nuclide&$T_{1/2}$&$Q_{\epsilon}$&$E_{\nu}$&$L_{23}/2$&$E_{e,max}$&weight&$\nu$- \\
&&(keV)&(keV)&(m)&(keV)&gr&intensity(s$^{-1}$)\\
\hline
\hline
$^{41}$Ca&$10^5$y&421&417&208&260&400&$10^{12}$\\
\hline
$^{51}$Cr&$28$d&753&747&373&560&250&$10^{18}$\\
\hline
$^{55}$Fe&$2.7$y&232&226&110&106&4000&$5\times10^{17}$\\
\hline
$^{71}$Ge&$11$d&232&222&110&100&300&$2\times10^{18}$\\
\hline
$^{103}$Pd&$17$d&543&480&240&315&15&$5\times10^{16}$\\
\hline
$^{109}$Cd&$460$d&214*&101&50&30&50&$5\times10^{15}$\\
\hline
 $^{139}$Ce&$138$y&113*&74&37&20&1.5&$2\times10^{14}$\\
 \hline
 $^{157}$Tb&70y&60.0(3)&9.8&5&0.4&5&$2\times10^{14}$\\
 \hline
$^{163}$Ho&4500y&$\approx 2.6$&$\approx 0.5$;$\approx 0.8$&0.2-1.3&$\le 0.03$&250&$5\times10^{12}$\\
&&&;$\approx 2.2$;$\approx 2.3$& & & &\\
& & & 2.6& & & &\\
\hline
 $^{193}$Pt&50y&568.0(3)&44($70\%$)&22&6.5&300&$5\times10^{14}$\\
 & & &54($30\%$)&27 & 9& &\\
\hline
\hline
\end{tabular}
\end{center}
\end{table}
\section{EXPERIMENTAl APPROACHES TOWARDS NEUTRINO OSCILLOMETRY}

One could propose different experimental methods, already mentioned above, which can be carried out for the scan of neutrino scattering events. 
One suggestion is to install the neutrino source in the centre of sphere and to measure the events in the $4\pi$-geometry which provides the highest angular efficiency, leadingg to the geometric factor $g_s=1$ as discussed above. This, however, is only suitable for relatively small $L_{32}$-values because the sphere dimensions are limited, at least very lsarge TPC  spheres are not presently available. Another method concerns to the use of long detector (cylinder type) with installation of the source on the top and/or the bottom of detector. Though angular efficiency is much smaller than in the case of sphere, $g_c\ll g_s$, the oscillation path could be much longer. In addition, the neutrino source can be removed each time, if necessary, in order to perform the background measurements or to install a new sample for repeated measurements.\\
In the calculations of the expected number of events for both these methods  one needs to know the neutrino intensity of a specific source (see Table \ref{tab1}) and the neutrino-electron elastic scattering cross-section.
  The dependence of this cross-section on neutrino energy is not sensitive to the elemental composition of the detector's target material and is shown in Fig. \ref{fig:sigmatotal} for the energy interval up to 750 keV.  

\begin{figure}[!ht]
 \begin{center}
 \includegraphics[scale=1.0]{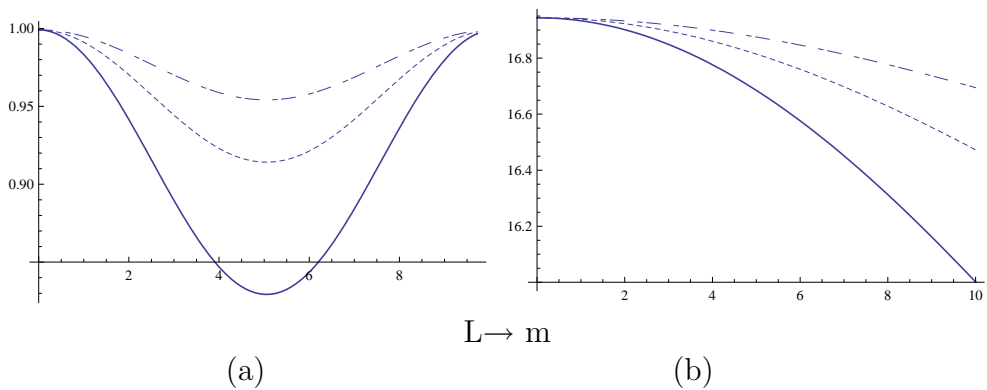}
 \caption{The  rate $\frac{dN}{dL}$ (per meter) for Ar at $10$ Atm with $1$ Kg of
 $^{157}$Tb (a) and $^{193}$Pt (b)
 as a function of
 the source-detector distance (in m ).
The results shown correspond to
$\sin^2{2\theta_{13}}=0.170,0.085$ and $0.045$ (decreasing from bottom to top). This rate was obtained for a running period equal to the half life of the source.
}
 \label{rates1and4_40}
  \end{center}
  \end{figure}
\subsection{Spherical gaseus TPC-detector with the internal monoenergetic neutrino source}
In this spherical chamber the neutrino source is proposed to be situated in the centre of the sphere and the electron detector is also placed around the source in the smaller sphere with radius $r=0.5$ m. The sphere volume out of the detector position is filled with a gas (a noble gas such as Ar or preferably Xe, which has a higher number of electrons). The recoil electrons are guided by the strong electrostatic field towards the Micromegas-detector \cite{Giomataris},\cite{GIOMVER08}. The NOSTOS, Time Projection Counter, has an advantage in precise position determination (better than 0.1 m) and in detection of very low electron recoils (down to a few hundred of eV, that suits to all nuclides from Table \ref{tab1} except of  $^{163}$Ho). 

We have seen that the number of electron events is given by Eq. \ref{eventsph}. This includes:
\begin{itemize}
\item The elastic neutrino-electron cross section. \\ 
 (see Eqs \ref{sigmatot1} and \ref{sigmatot2}). The scale is set by the value given in Eq. \ref{weekval}
 \item The number of neutrinos emmitted by 1 Kg of the source.\\
\beq
N_{\nu}=\frac{m_s}{1~Kg} \frac{1}{A_s \times 1.66\times 10^{-27}}=\frac{m_s}{1~Kg}\frac{6.0}{A_s}\times 10^{26}
\eeq
where $A_s$ is the atomic number of the source and $m_s$ is its mass.
 For any given time $t$ the number of neutrinos emitted must be multiplied by the fraction
$1-e^{-t/\tau}$ with $\tau=\frac{T_{1/2}}{\ell n 2})$.
\item The number of electrons present in the target:\\
Assuming that we have a gas target under pressure $P$ and temperature $T_0$ we find:
\beq
n_e= Z\frac{P}{kT_0}=4.4\times 10^{27}m^{-3} \frac{P}{10~Atm}\frac{Z}{18}\frac{300}{T_0}
\eeq
($Z$ the atomic number).
\end{itemize}
%

By using  this value, as well as the neutrino intensity and the neutrino-electron cross section (see Fig. \ref{fig:sigmatotal}), we estimated the expected total neutrino scattering events as a function of distance from the centre of sphere, for sources $^{193}$Pt and $^{157}$Tb, including the oscillation effect. Here we assumed that the sphere radius was equal to r=10 m and the Ar target was under pressure P=10 Atm  at room temperature. We have considered three values of the parameter $sin^2 (2\theta_{13})$, namely  $sin^2 (2\theta_{13})= 0.170,0.085,0.045$. These values were chosen in accordance with the  estimate of  $sin^2\theta_{13} = 0.02\pm0.01$ recently obtained \cite{FOGLI} .  
 
 The obtained results are shown
in Figs \ref{rates1and4_40}a and \ref{rates1and4_40}b. In these plots one should pay attention not only to the absolute value of the rate but also to the dip due to oscillation. Clearly very accurate measuremets are needed. From the length of oscillation point of view the ideal choice is $^{157}$Tb, since the full oscillation takes place inside the detector, but $^{193}$Pt could be a good compromise.

\subsection{Cylindrical detector with the external monoenergetic neutrino source }

An alternative approach consists in using  a long cylinder filled with a liquid target, which can provide much longer scanning length. Since the liquid  target has higher density than a the gaseous one of the spherical TPC, one expects higher event rates. Admittedly, however,  for an external target installed on the top of the detector, there will be flux losses caused by smaller effective acceptance angle, if the whole range of length $L$ is to be explored. This will be espacially true for large $L$, in which case the cone with the vertex at the place of source will be smaller than a half sphere, let alone a full sphere.  As a matter of fact we have seen  that the geometric factor, compared to that of a half sphere, is smaller (see Fig. \ref{fig:geom}). These  losses are in some sense compensated by the possibility to scan the long $L_{32}$-values which are attributed to relatively short-lived nuclides, which results to higher neutrino source intensities that can be produced (see Table \ref{tab1}). 

The use of an external and reproducible source has very big advantage because allows to measure the background when source is taken away. This background for the  proposed experiment can be generated by cosmic and partly by geo-neutrinos. 
As an example one can estimate the neutrino scattering rates for the cylindrical detector LENA \cite{OBERAUER} which can be filled by liquid scintillator in the tank with the diameter 30 m and height 90 m. The parameters of this detector allow to reach the expected position resolution of ca. 60 cm and the energy resolution of 40 keV for 200 keV energy.       
The best cosmic background conditions are reached at 700 keV energy for the LENA-detector. Therefore in Table \ref{tab1} we include the $^{51}Cr$ with the neutrino energy 747 keV with a correspondingly long oscillation length in comparison to the length of detector (90 m). One may be able to cope with this problem by satisfying oneself with only a portion of the oscillation inside the detector. The strong source of  $^{51}$Cr, of at least 30 MCi,  can be prepared for this case. 
Figs \ref{cyzFe2}- \ref{cyzPd2} show estimated total rates for the electron-neutrino scattering for sources $^{51}$Cr, $^{55}$Fe, $^{71}$Ge and $^{103}$Pd for the above values of $sin^2(2\theta_{13})$ . In estimating the value of $\Lambda$ the electron density of $n_e= 2\times 10^{29}$m$^{-3}$ for LS-LAB will be used. 

 In this case the presentation of the results is more complicated, since the total rate, for a given neutrino energy, depends on two variables $\rho$ and $\zeta$ (see
  Eq. (\ref{Eq:cylinder})) and the geomtric factor is a bit complicated. We will present our results in units of $\Lambda=\frac{G^2_F m^2_e}{2 \pi}N_{\nu} n_e R $  (see Eq. (\ref{gcLambda})). Thus only the dependence on the neutrino energy and the geometry will be exhibited.
  
  Naively one might expect to see the oscillation patter in the case of cylindrical geometry directly from the raw data in a fashion analogous to that of the sphere. This naive view, however, is not true as seen in Fig. \ref{cyFerz1}. The variation of the neutrino flux from point to point outweighs the variation of the oscillation amplitude from point to point. Clearly a much more careful analysis should be employed. This is indeed achieved by counteracting the effects of the flux variation, i.e. by multipling the event rate by $(\zeta^2+u^2 \rho^2)$. This way we obtain results as follows:
 \begin{figure}[!ht]
 \begin{center}
\includegraphics[scale=1.0]{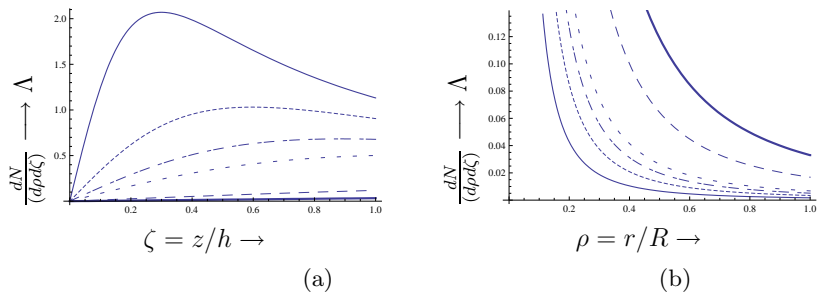}\\
\hspace*{2.0cm} { (a)}
\hspace*{4.0cm} { (b)}\\
 \caption{The  differential event rate $\frac{dN}{d \rho d \zeta}$ , in units of $\Lambda$ (see text), for a cylinder of radius $r=15$m and length $h=90$ m 
  and $E_{\nu}=226$ keV. On the left panel (a) the plots are preseted  as a function of $\zeta=z/h$ for the values of  $\rho=r/R=(0.05, 0.1, 0.15, 0.2, 0.5, 1.0)$,  to be read from top to bottom. On the right (b) the same quatity is presented as a function of $\rho=r/R$ for the values of  $\zeta=z/h=(0.05, 0.1, 0.15, 0.2, 0.5, 1.0)$. The effect of oscillation is not visible. It is included only for orientation purposes, i.e. to demonstrate that  a more careful analysis is needed.
}
 \label{cyFerz1}
  \end{center}
  \end{figure}
   \begin{itemize}
 \item Rates for the source  $^{55}$Fe (neutrino energy 226 keV):\\
 In this case from the data of table \ref{tab1} and the above value of $n_e$ we obtain $\Lambda=0.243$s$^{-1}=7.7\times 10^6$y$^{-1}$.\\
 If we multiply the obtained event rate by the above mentioned factor, $(\zeta^2+u^2 \rho^2)$,  we get the results shown in Fig. \ref{cyzFe2}- \ref{cyzPd2}. 
  From Fig. \ref{cyzFe2} we see that the magnitude of the rate depends strogly on the radial distance from the axis, but the oscillation period depends only mildly on it.
  \begin{figure}[!ht]
 \begin{center}
 \includegraphics[scale=1.0]{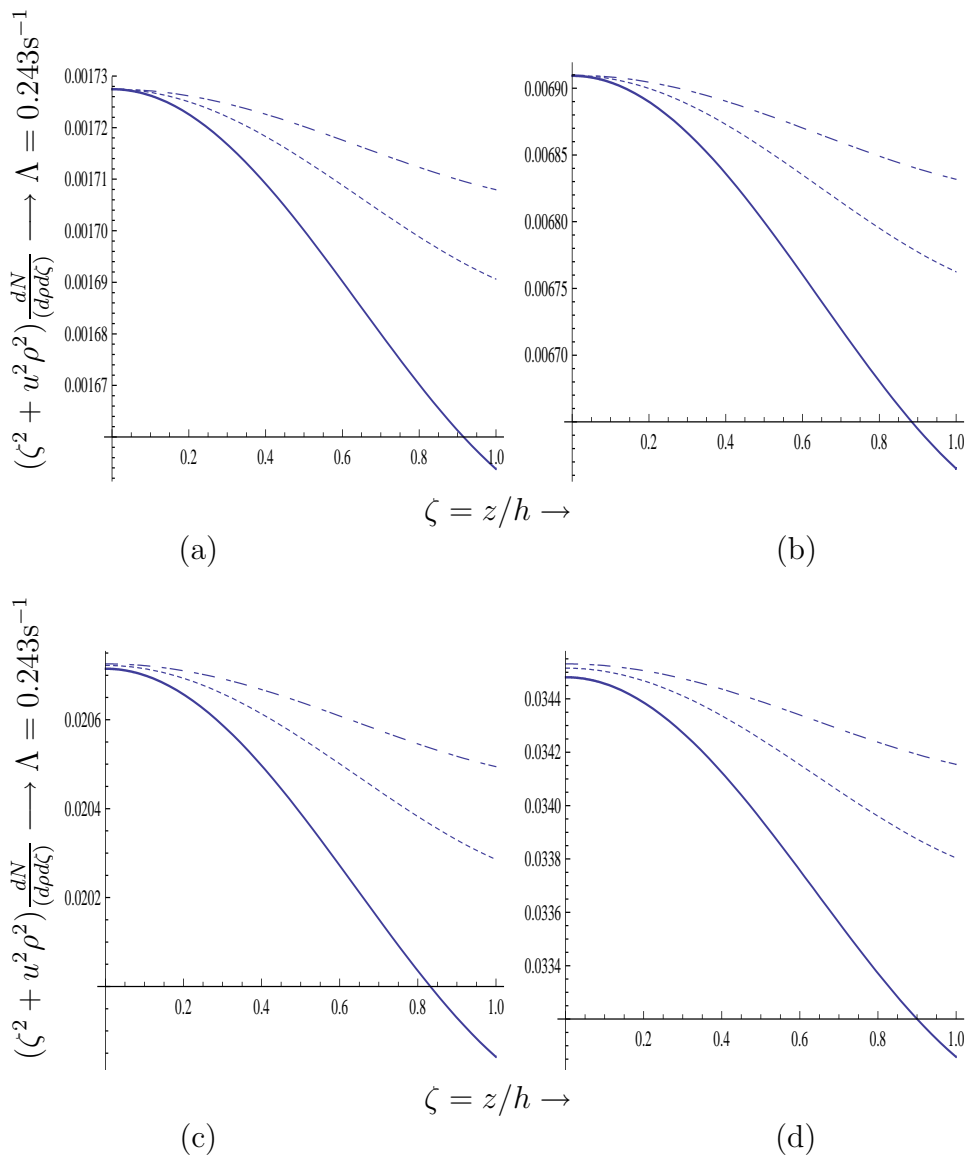}
 \caption{The  differential event rate $(u^2 \rho^2+\zeta^2)\frac{dN}{d \rho d \zeta}$ (in units of $\Lambda$) for a cylinder of Radius $R=15$m and length $h=90$m as a function of $\zeta=z/h$ for $^{55}$Fe ($E_{\nu}=226$ keV). The solid, dotted and dashed curves correspond to $\sin^2{2 \theta_{13}}$= 0.170, 0.085 and 0.045 respectively. The panels (a), (b), (c), (d) 
  correspond to values of $\rho=r/R$ =0.05, 0.2, 0.6 and 0.8 respectively.  
}
 \label{cyzFe2}
  \end{center}
  \end{figure}
  
 \item Rates for the source  $^{51}$Cr (neutrino energy 747 keV).\\
 In this case $\Lambda=0.577$s$^{-1}=1.82\times 10^7$y$^{-1}$
 
   \begin{figure}[!ht]
 \begin{center}
  \includegraphics[scale=1.0]{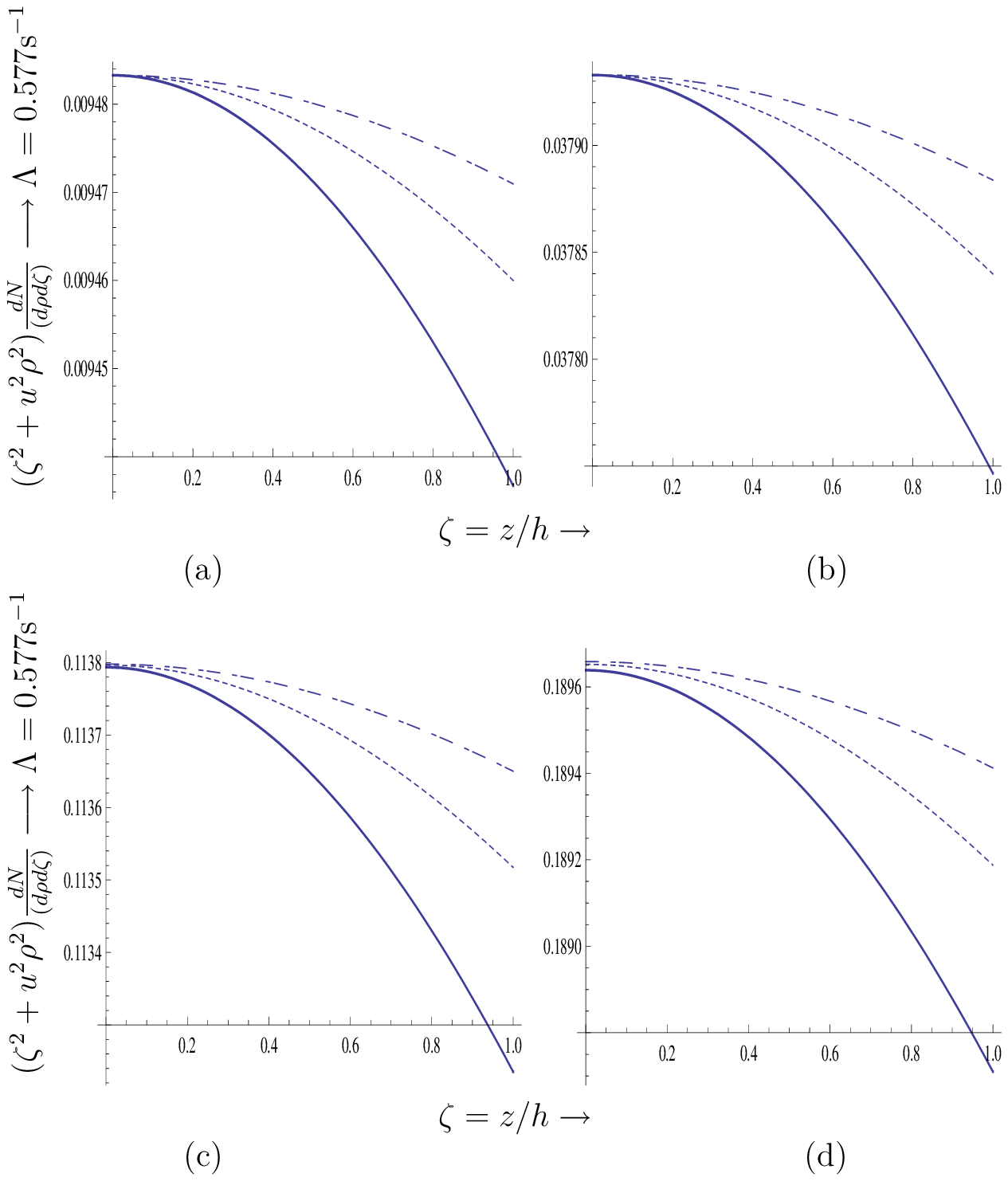}
 \caption{The same as in Fig. \ref {cyzFe2}  for $^{51}$Cr ($E_{\nu}=747$ keV).
}
 \label{cyzCr2}
  \end{center}
  \end{figure}  
  \item Rates for the source $^{71}$Ge (neutrino energy 222 keV).\\ 
  In this case $\Lambda=1.266$s$^{-1}=4.0\times 10^7$y$^{-1}$
   \begin{figure}[!ht]
 \begin{center}
  \includegraphics[scale=1.0]{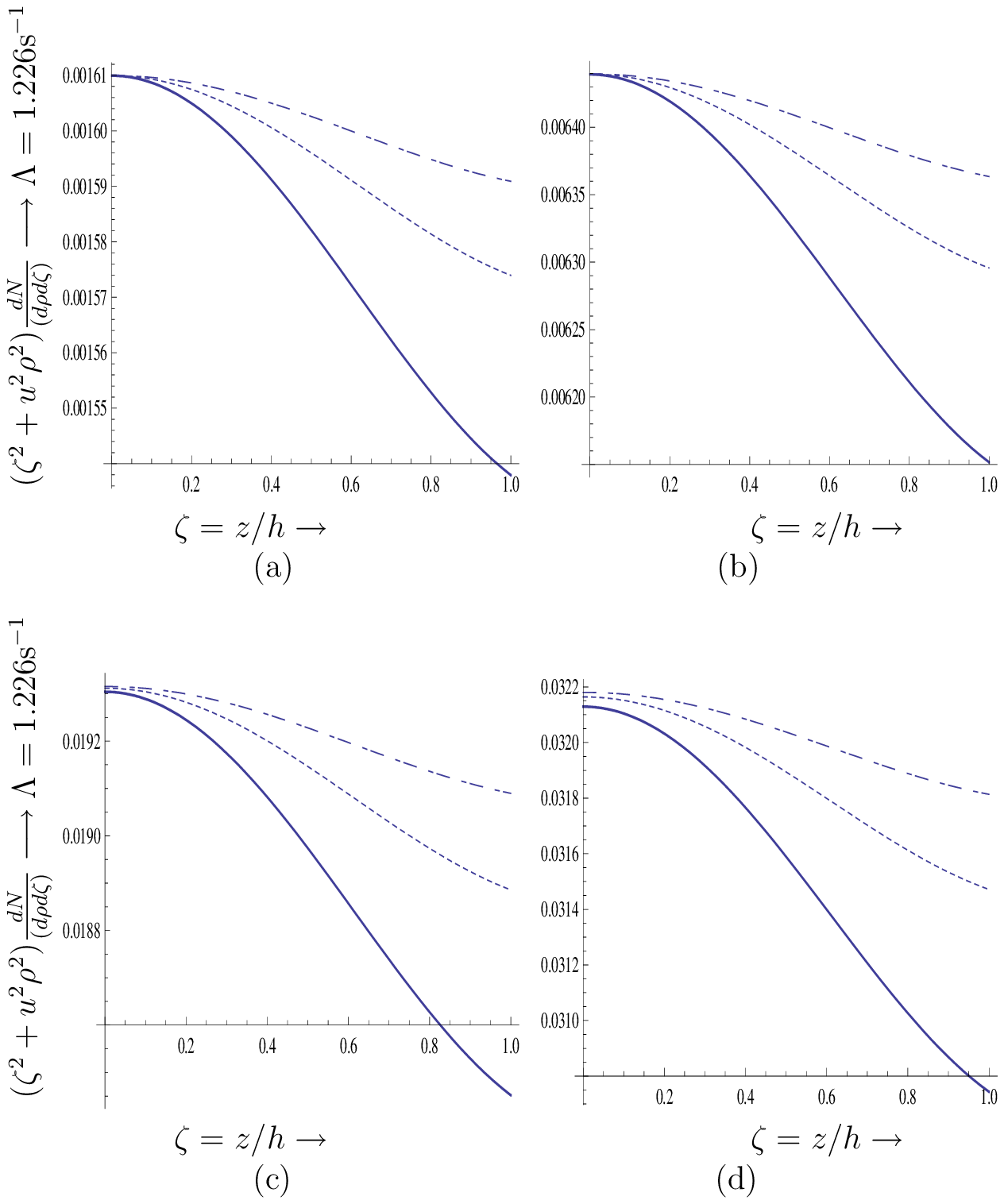}
 \caption{The same as in Fig. \ref {cyzFe2}  for $^{71}$Ge ($E_{\nu}=222$ keV).
}
 \label{cyzGe2}
  \end{center}
  \end{figure}  
  \item Rates for the source $^{103}$Pd (neutrino energy 480 keV).\\
  In this instance $\Lambda=0.0282$s$^{-1}=891$y$^{-1}$
  
   \begin{figure}[!ht]
 \begin{center}
   \includegraphics[scale=1.0]{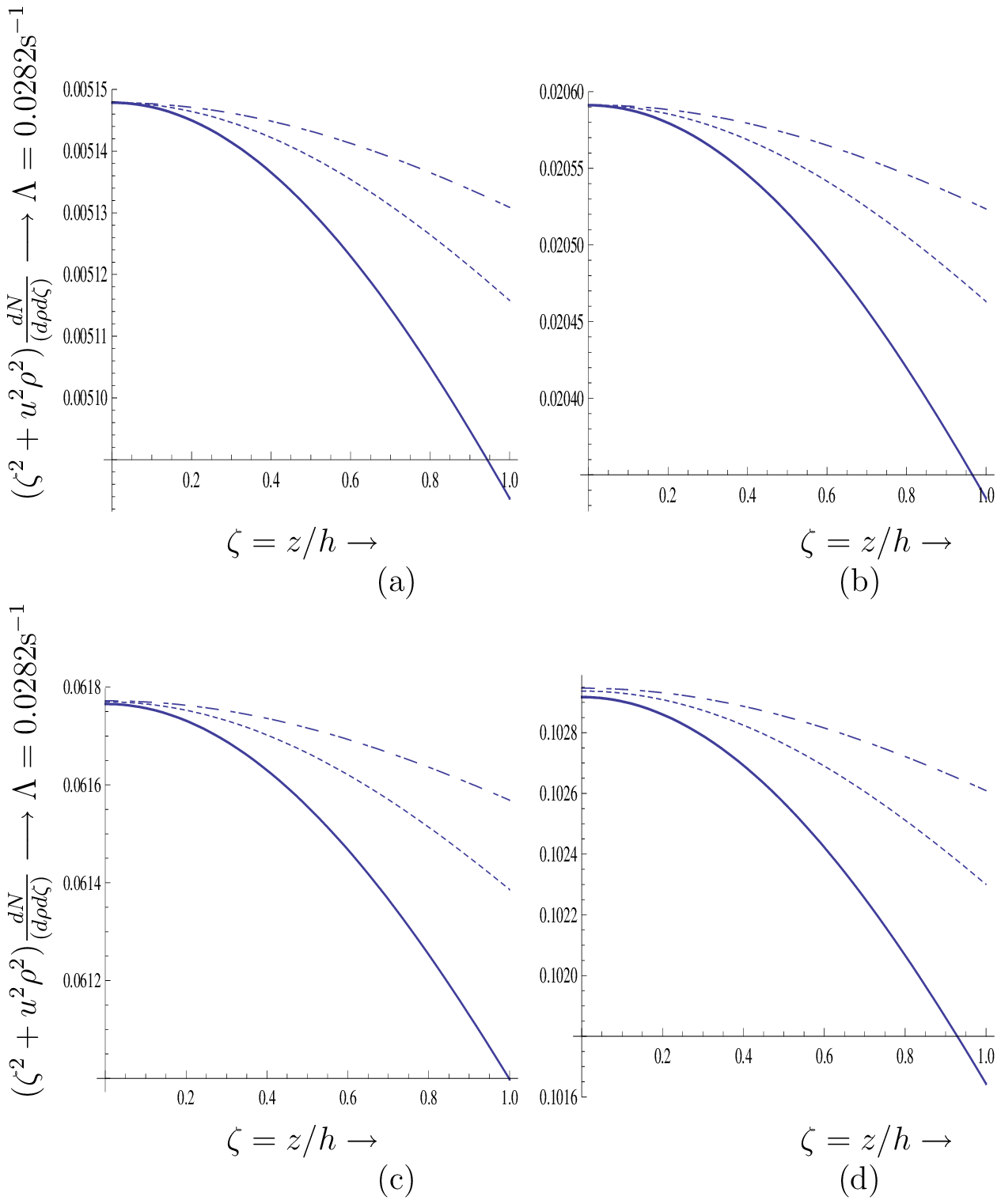}
 \caption{The same as in Fig. \ref {cyzFe2}  for $^{103}$Pd ($E_{\nu}=480$ keV).
}
 \label{cyzPd2}
  \end{center}
  \end{figure}  
 \end{itemize}
 It is rewarding that, if the data analyzed as above, the oscillation pattern exhibited by a cylindrical detector  is not very different from that of the spherical detector. It should be kept in mind that the exhibited range of $\zeta$ is to guide the eye. The maximum value of $\zeta$ is, by definition, unity.
 \subsection{Analysis of the data of a cylinrical detector in terms of $L$}
\begin{figure}[!ht]
 \begin{center}
   \includegraphics[scale=1.0]{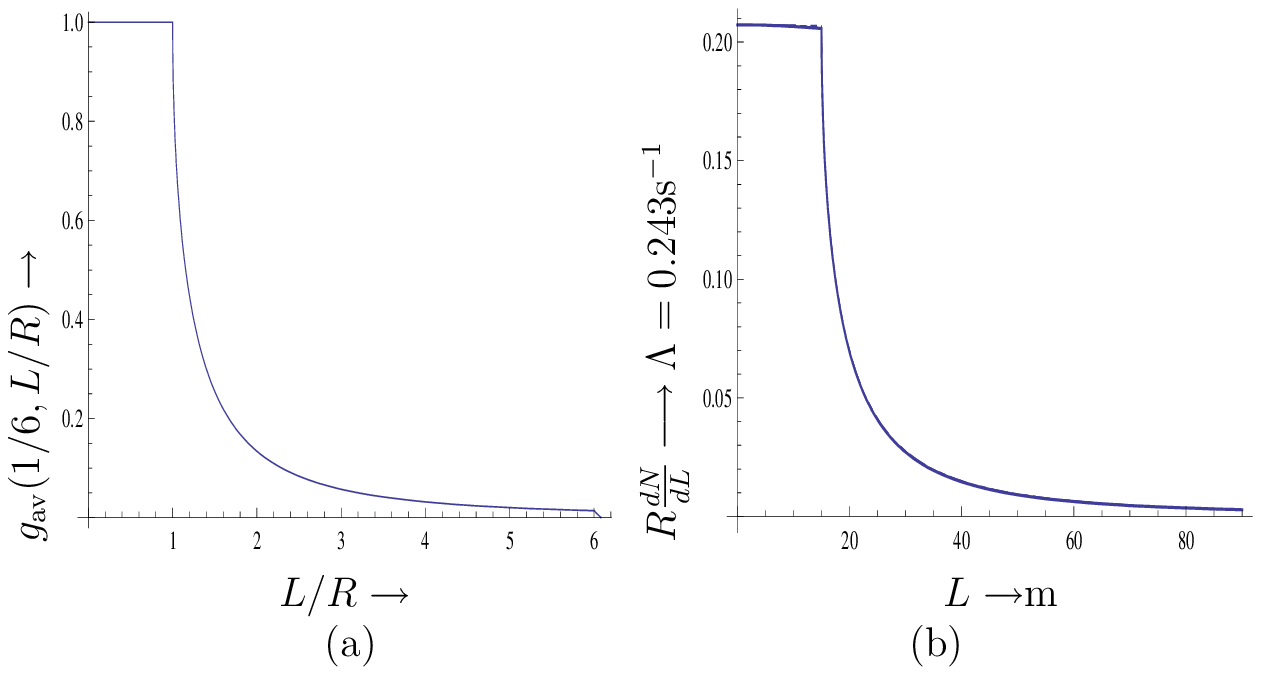}
 \caption{In panel (a) we show the geometric factor $g_{\text{av}}(1/6,L/R)$ as a function of $L/R$ for the value $u=R/h=1/6$. Note the flux factor of 1/2 has explicitly put in in Eq. (\ref{Eq:cyrav}) and thus it is not included in the geometry. In the first region, which is a spherical segment,  one sees that the geometric factor is unity like the case of the sphere. In panel (b) we show the expected event rate (in units $\Lambda=0.243$s$^{-1}$) in the case of the $^{55}$Fe source as a function of L in meters. It is obvious that both plots follow the same pattern.}
 \label{cygav}
  \end{center}
  \end{figure}

\begin{figure}[!ht]
 \begin{center}
  \includegraphics[scale=1.0]{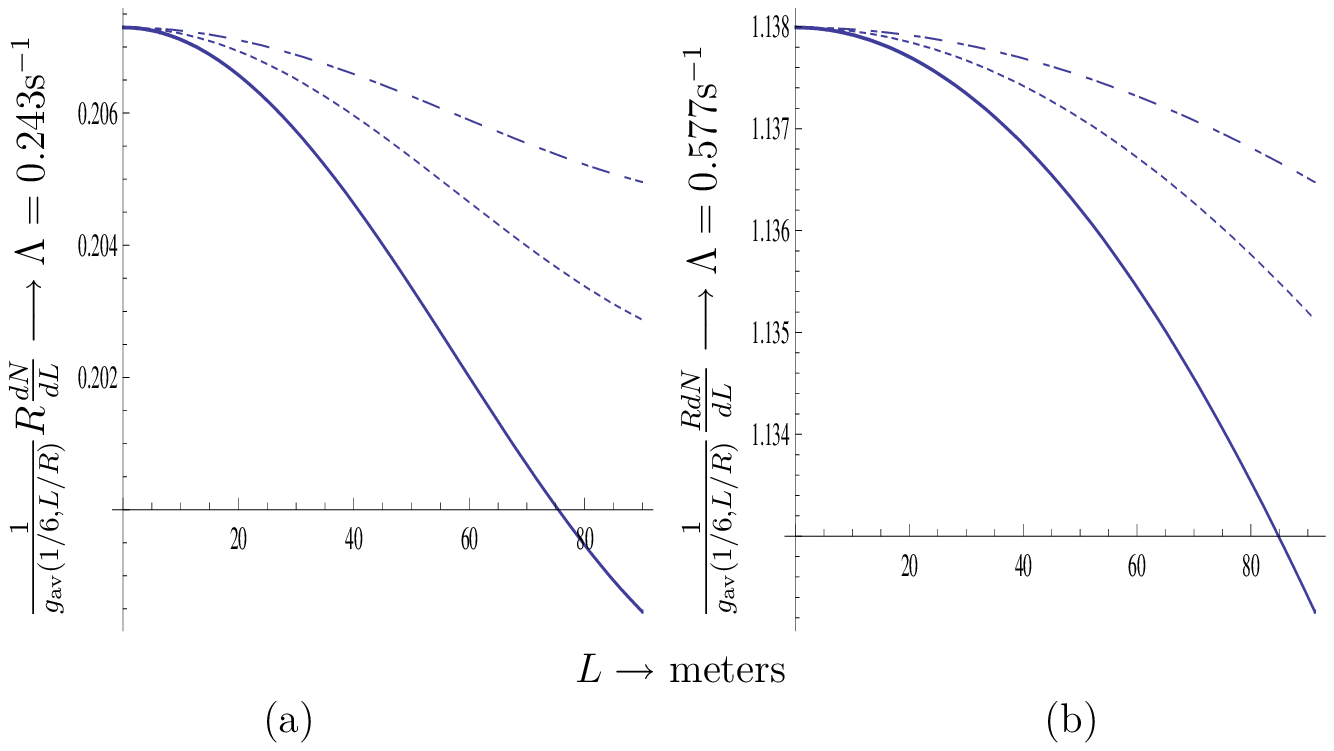}
 \caption{In panel (a) we show the expected event rate $R\frac{dN}{dL}$, divided by the geometric factor, in units of $\Lambda=0.243$s$^{-1}$ expected for the source $^{55}$Fe, while in panel (b) we show the same quantity for $^{51}$Cr in units of $\Lambda=0.577$s$^{-1}$. In both panels the solid, dotted and dashed curves correspond to $\sin^2{2 \theta_{13}}$= 0.170, 0.085 and 0.045 respectively. One does not see the full oscillation inside the detector, but the information is adequate to extract usefull information on the neutrino oscillation parameters.}  
 \label{rate.aveFeCr}
  \end{center}
  \end{figure}

\begin{figure}[!ht]
 \begin{center}
  \includegraphics[scale=1.0]{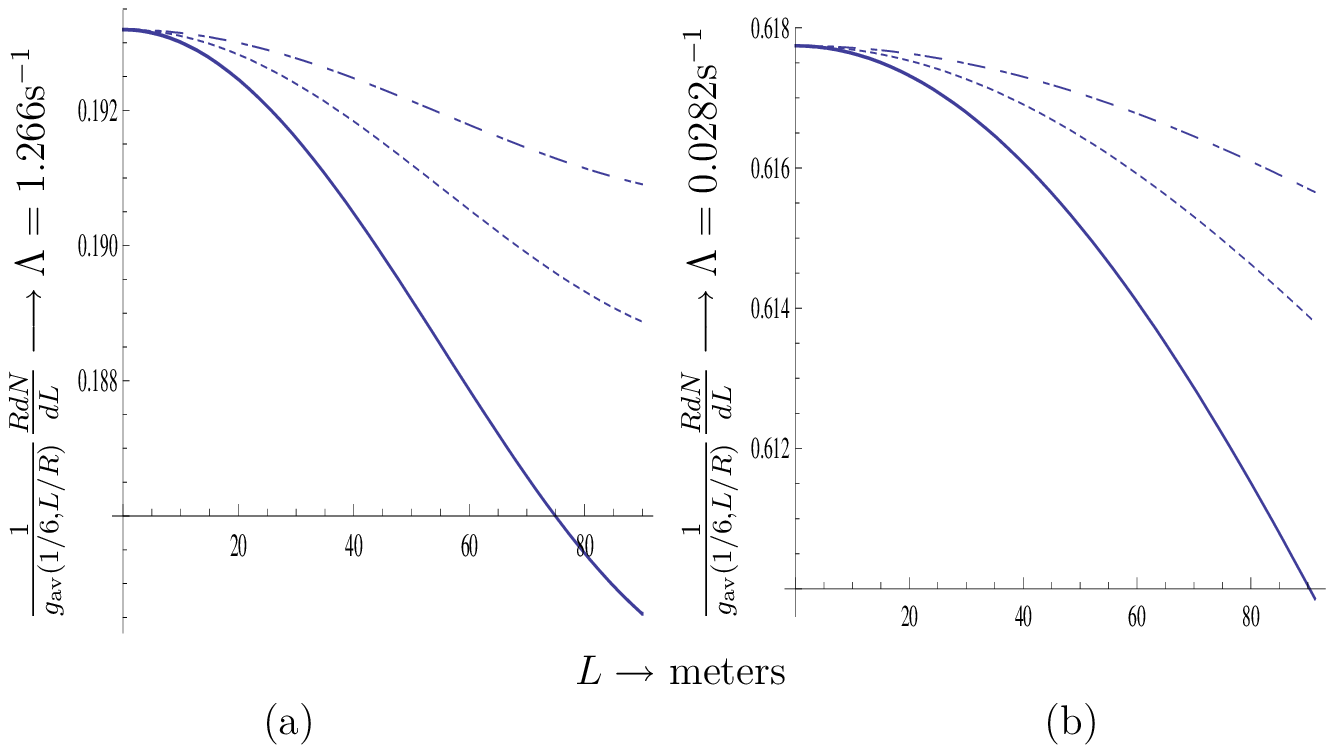}
 \caption{The same as in Fig. \ref{rate.aveFeCr} in case of the source $^{71}$Ge (a) and $^{103}$Pd (b).}
 \label{rate.aveGePd}
  \end{center}
  \end{figure}

 In the previous section we analyzed the data by obtaining the event rate for each point $\rho,\zeta$ of the detector. One may attempt to present the data in terms of the source detector distance $L$ by a suitable transformation from the variables $(r,z)$ to $(L,\phi)$ and integrating over the angle $\phi$. A tedious but straightforward algebra yields:
 \beq
 R\frac{dN}{dL}=N_{\nu} n_e R\frac{1}{2} g_{\text{av}}(u,L/R)\sigma(E_{\nu},L)=\Lambda\frac{1}{2} g_{\text{av}}(u,L/R) \tilde{\sigma}(E_{\nu},L)
 \label{Eq:cyrav}
 \eeq
 where $g_{\text{av}}(u,L/R)$ is a geometric factor that takes care of the variation of the neutrino flux in the various positions described by $L$. It can be cast in the form:
 \beq
 \left \{
 g_{\text{av}}(u,\upsilon)=\begin{alignedat}{2}
 &1,\quad&0<\upsilon<1\\&1-\sqrt{\upsilon^2-1}/\upsilon,\quad& 1<\upsilon<1/u\\&(1/u-\sqrt{\upsilon^2-1})/\upsilon,\quad& 1/u<\upsilon<\sqrt{1+1/u^2}
 \end{alignedat}
 \right.
 \label{Eq:gav}
 \eeq
 This funcion for $u=1/6$ is shown in Fig. \ref{cygav}. On the same figure we show the expected event rate for a typical case like $^{55}$Fe.
  It is clear that to disentangle the oscillation dependence on $L$ one must divide the expected rate by the geometric factor. When this is done we obtain the results shown in Figs \ref{rate.aveFeCr} and \ref{rate.aveGePd}.
   
 \section{Discussion}
 In this article we consider two possible experimental approaches for the neutrino oscillometry: the spherical gaseous detector and the liquid scintillator cylindrical one. 
Both cases have their own advantages and disadvantages. 
\begin{itemize}
\item The spherical gaseous detector.\\
This requires the construction of a sphere with the radius much larger than 10 meters  can hardly be built. This restricts the possible candidates for the neutrino oscillometry by nuclides whose neutrinos  energies  should  not be much higher than 10 keV in accordance with the equation \ref{L32}. The relevant oscillometry curves are shown in Fig. \ref{rates1and4_40} employing an Ar gas.
One  kg of source, installed at the center of the sphere, was employed under the  assumption of a time measurement longer than the half-life of the source.
To obtain the number of events collected during the reasonable measurement time, the values in Figs \ref{rates1and4_40} must be multiplied by a number of disintegrated nuclides during this time. Thus, e.g., for one year measurement with the 0.15 kg of $^{157}$Tb one can obtain the rate equal approximately to $8\times 10^{-4 }$events per meter, too low value. 
The similar feature is expected for other relatively weak neutrino emitters listed in the bottom part of Table \ref{tab1}.  Thus we can conclude that the spherical gaseous detector with the neutrino source in the center, being very attractive in its idea, cannot provide presently the appropriate conditions for neutrino oscillometry. As a matter of fact, the requirement of low energy neutrinos brings the long-lived candidates with the low rate of events. The small neutrino energy is associated with the tiny neutrino-electron cross-section, and the small electron recoil energy requires the use of gaseous environment with the TPC-detection  which has not a high target electron density. All these disadvantages with an additional demands for very long irradiation at the reactor for the production of appropriate nuclides and a very long data acquisition time makes this approach very problematic. 
\item A long cylindrical detector.\\
 In this case  the neutrino source in a form of a small sphere can be installed  at the top/bottom of the cylinder at the center of the  circular face.  Even though the angular efficiency of such configuration is much less than for $4\pi$ acceptance of the spherical detector,  there are many advantages which overbalance this lack. If the length of cylinder is rather long (let's presume about 100 m) the neutrino emitters with the energies in the hundreds keV region can be used for oscillometry. This on one hand increases the neutrino-electron cross-section  (see Fig.  \ref{fig:sigmatotal}), while on the other  drastically increases the neutrino source decay rate because of decreasing the source half-life. Then the increased in the energy of the recoiling electrons  makes possible to use a liquid target, which leads to the increase of the number of target electrons.
 
As it can be seen from the top part of Table \ref{tab1} there are several candidates ($^{51}$Cr, $^{55}$Fe, $^{71}$Ge, and $^{109}$Pd) for whom the neutrino flux is a few order of magnitudes higher than for the heavier nuclides.  The $(\nu_e,e)$ cross-section for, e.g. $^{71}$Ge or $^{55}$Fe, are at least  two orders of magnitude higher than that for $^{157}$Tb, a typical candidate for oscillometric measurements in the spherical gaseous TPC. One-two orders of magnitude can be won for electron density in the liquid target. Thus, approximately eight orders of magnitude gain for a long cylindrical liquid detector is against about of one-two orders of magnitude angular efficiency loss in comparison to spherical gaseous case. Meanwhile, there is some disadvantage in the use of cylinder whose length is restricted by 100 m or a little bit more. For relevant candidates collected in the \ref{tab1} with the neutrino energy higher than 220 keV the scan in a full oscillation length (including both disappearance and appearance events) is impossible. Meanwhile, as it can be seen from Fig. \ref{cyzFe2}, the oscillation curves can be well recognized (till z/h =1) and the dependence on the  mixing angle $\theta_{13}$  is also visible.     
\end{itemize}  
 
In order to show the feasibility of neutrino oscillometry for long cylindrical liquid target let's comment the curves in Fig. \ref{cyzFe2}. These curves have been calculated for the neutrino energy of 226 keV in the cylinder with the length $h=90 m$ and with the radius $R=15 m$ filled by the liquid scintillator (LS).  The scale parameter $\Lambda= 0.243 $s$^{-1}$ appearing in these  curves  can be obtained with  assumption that the neutrino source intensity is  $N_{\nu} =5 \times 10^{17}$ s$^{-1}$ (see Table \ref{tab1}) and  an electron density $2\times 10^{29}$m$^{-3}$ via Eq. \ref{gcLambda}. As a factor $u^2 \rho ^2+\eta ^2=(r^2+z^2)/h^2$ it is less than 1 for practically full neutrino path in the cylinder, the range of values of the  average differential rate,  ${\partial^2 N}/{\partial \rho \partial \zeta}$, is from $>10^{-3}$ s$^{-1}$- for the top left curves of Fig. \ref{cyzFe2} to $>2\times 10^{-2}$s$^{-1}$ for the bottom right ones. The values of differential accounts strongly depend on the $(r,z)$-position of event in the cylinder.
Another evidence for successful oscillometry can be shown from the presentation of the data in the   ${\partial^2 N}/{\partial \rho\partial \zeta}$ -values. Though the expected position resolution in the cylinder with LS is better \cite{OBERAUER} than 1 m , we will assume the volume resolution equal to 1 m$^3$. 
Fig. \ref{cyzFe2} shows an experimental "fingerprint" which is expected from the existence of neutrino oscillations.
  
 The solar neutrino background depends on the energy region of interest and it becomes small above 400 keV.The expected background from the solar events in this region  is $< 1$ day$^{-1}$m$^{-3}$ for an LS detector of the size used in our calculations. It is thus useful to obtain some oscillation features in this energy region, using the two sources $^{51}$Cr and $^{103}$Pd, which are favorable in this region. The oscillation features of these nuclides are shown in Figs \ref{cyzCr2} and \ref{cyzPd2}. One can obtain at the end of the cylinder a difference of $8.7\times 10^{-3}R$s$^{-1}$ for values $\sin^2{2 \theta_{13}}=0.170$ and $0.045$ in the case of $^{51}$Cr. This yields $5.8\times10^{-4}$m$^{-1}$s$^{-1}$. Taking into account the radioactive decay of $^{51}$Cr we obtain 1560 events per meter for 60 days of measurent. This difference exceeds the 4$\sigma$ standard statistical uncertainty, even for this unfovarable candidate. Note that the estimated rates in the liquid scintillator presented in Fig. \ref{cyzCr2} are higher than expected background mentioned above.

More impressive results are expected in the case of $^{55}$Fe and $^{71}$Ge, which are characterized by the fact that almost the full oscillation takes place inside the detector (see Figs \ref{cyzFe2} and \ref{cyzGe2} as well as \ref{rate.aveFeCr} and \ref{rate.aveGePd}). We should mention, however, that in this energy one encounters higher solar neutrino background\cite{BU-RMP88}.

 Quite high values obtained can be even increased if the data handling is going during several months or even weeks. These measurements can be stopped in any time. Source can be moved and new source installed. Because of high neutrino intensity the reasonable result can be obtained very quickly without the agonizing long-term measurements. This advantage of removable sources is indeed very  attractive. 

\section{CONCLUSIONS}

 The new method for neutrino oscillation measurements is proposed. It is based on the use of  monoenergetic low energy neutrinos which are released in the atomic electron capture by nuclei. High intensity neutrino sources with the neutrino energies less than a few hundreds keV can be produced in neutron reactors. They can  be installed at the top of the long cylinder filled by liquid target material or placed in the center of gaseous sphere placed underground. 
The measurement consists in finding  the number of events arising from the electron neutrino scattering on the electrons of the target. It was shown that, for some appropriate candidates like ($^{55}Fe, ^{71}$Ge, $^{109}$Cd),  the so called small oscillation length $L_{32}$ can fall within the dimensions of cylindrical detectors which can be constructed in the nearest future. There the oscillations can be determined by disappearance of events caused by the change of the neutrino flavor just inside the detector. These events can be scanned point-by-point thus providing the oscillometry curve. The calculations for the cylindrical detector with the length $h=90$ m and radius $R=15$ m showed a complicated net of events which can be analyzed by the  "oscillation signature" suggested.  The most attractive advantage of proposed experiment is the possibility to move the neutrino source, to measure the background or to implement other physics project and also to replace one source by another. With the neutrino source intensity of 30 MCi, which provides high count/background rate, the observation of oscillations can be performed during months or even weeks in underground conditions .        
Since the $\theta_{13}$ discovery potential for proposed experiments spans the values of $sin^2 2\theta _{13} <0.170$, the  experiments with low monoenergetic neutrinos, e.g., with the LENA-detector, could be complementary to investigations with the reactor and accelerator neutrinos in Double Chooz, RENO, Daya Bay, T2K and NOvA projects planned for the coming decade.    
The physics interest in the short baseline oscillometry measurements is not only the determination of  $\theta_{13}$. The discovery of the $L_{32}$ -value from the oscillometry experiments with the subsequent comparison with the electron neutrino energy will allow to check availability of equation (\ref{L32}), the dynamics of $\nu_e-e $ low energy interactions and, in general, the validity of the long baseline oscillation results obtained so far.  The present analysis does not suffer from the 8-fold degeneracy \cite{BMW02} that corrupts the measurement of $\theta_{13}$ in conventional long-baseline experiments. The neutrino energies, needed to check (\ref{L32}), can be accurately determined by independent direct mass measurements of nuclides, a capture-partners, in the ion traps \cite{BNW10}.

acknowledgments: This work was supported in part by the
European Union under the contract
MRTN-CT-2004-503369 and by the program PYTHAGORAS-1 of the
Operational Program for Education and Initial Vocational Training of the
Hellenic Ministry of Education under the 3rd Community Support Framework
and the European Social Fund. Support by the Russian ministry of science is also acknowledged.\\
 We would also like to thank Y. Giomataris and A Vasiliev for their very fruitful remarks and discussions. Fruitful discussions during the preparation of this work   with Ya. Azimov, E. Akhmedov, K. Blaum, S. Eliseev, T. Enqvist, A. Erikalov, F. von Feilitzsch, W. Hampel, V. Isakov, H.-J. Kluge, U. Koester, M. Lindner, K. Loo, A. Merle, T. Oberauer, W. Trzaska, J. Winter, M. Wurm, and A. A. Vorobyov  are also happily acknowledged. 


\end{document}